\documentclass{article}
\usepackage[english]{babel}
\usepackage[utf8]{inputenc}
\usepackage{base}
\usepackage{amsmath}
\usepackage{booktabs}
\usepackage{multirow}
\usepackage{caption}
\usepackage{subcaption}
\usepackage[T1]{fontenc}
\usepackage{xcolor}
\newcommand{\rebutal}[1]{{\color{black}#1}}
\usepackage{algorithm}
\usepackage{algpseudocode}
\usepackage{tcolorbox}
\definecolor{darkerlogocolor}{RGB}{20, 0, 145}  

\newtcolorbox{ttcolorbox}[1][]{colframe=darkerlogocolor, colback=darkerlogocolor!4!white, title=#1}

\newtcolorbox{apxtcolorbox}[1][]{colframe=black, colback=black!3!white, title=#1}

\title{Automated radiotherapy treatment planning guided by GPT-4Vision}

\author{Sheng Liu$^{1,2}$\footnote{Co-first authors}, Oscar Pastor-Serrano$^{1}$$^{*}$, Yizheng Chen$^{1}$, Matthew Gopaulchan$^{1}$,\\
Weixing Liang$^{3}$, Mark Buyyounouski$^1$, Erqi Pollom$^1$, Quynh-Thu Le$^1$,\\ Michael Gensheimer$^1$, Peng Dong$^{1}$, Yong Yang$^{1}$, James Zou$^{2,3}$\footnote{Correspondence: jamesz@stanford.edu, lei@stanford.edu}, and Lei Xing$^{1}$$^{\dagger}$ \\
        \small $^{1}$Department of Radiation Oncology, Stanford University, Stanford, CA, USA \\
        \small $^{2}$Department of Biomedical Data Science, Stanford University, Stanford, CA, USA \\
         \small $^{3}$Department of Computer Science, Stanford University, Stanford, CA, USA \\
         \\
}
\date{}

\begin{document}
\maketitle
\begin{abstract} 
\textbf{Objective}: Radiotherapy treatment planning is a time-consuming and potentially subjective process that requires the iterative adjustment of model parameters to balance multiple conflicting objectives. Recent advancements in frontier \rebutal{Artificial Intelligence (AI)} models offer promising avenues for addressing the challenges in planning and clinical decision-making. This study introduces GPT-RadPlan, an automated treatment planning framework that integrates radiation oncology knowledge with the reasoning capabilities of large multi-modal models, such as GPT-4Vision (GPT-4V) from OpenAI. 

\textbf{Approach}: Via in-context learning, we incorporate clinical requirements and a few \rebutal{(3 in our experiments)} approved clinical plans with their optimization settings, enabling GPT-4V to acquire treatment planning domain knowledge. The resulting GPT-RadPlan system is integrated into our in-house inverse treatment planning system through an \rebutal{application programming interface (API)}. For a given patient, GPT-RadPlan acts as both plan evaluator and planner, first assessing dose distributions and dose-volume histograms (DVHs), and then providing ``textual feedback'' on how to improve the plan to match the physician's requirements. In this manner, \rebutal{GPT-RadPlan} iteratively refines the plan by adjusting planning parameters, such as weights and \rebutal{dose objectives}, based on its suggestions. 

\textbf{Main results}: The efficacy of the automated planning system is showcased across \rebutal{17 prostate cancer and 13 head \& neck cancer VMAT plans with prescribed doses of 70.2 Gy and 72 Gy, respectively}, where we compared GPT-RadPlan results to clinical plans produced by human experts. In all cases, GPT-RadPlan either outperformed or matched the clinical plans, demonstrating superior target coverage and reducing organ-at-risk doses by \rebutal{5 Gy on average (15\% for prostate and 10-15\% for head \& neck)}. 

\textbf{Significance}: Consistently satisfying the \rebutal{the dose-volume objectives} in the clinical protocol, GPT-RadPlan represents the first multimodal large language model agent that mimics the behaviors of human planners in radiation oncology clinics, achieving promising results in automating the treatment planning process without the need for additional training.

\end{abstract}

\keywords{radiation therapy, treatment planning, large language models; multi-modal large language models; deep learning; artificial intelligence}\\

\section{Introduction}
Treatment planning of modern radiation therapy (RT) modalities, such as intensity modulated RT (IMRT) and volumetric modulated arc therapy (VMAT) is an inverse process solving an optimization problem based on dose prescriptions \cite{ oelfkeInversePlanningPhoton2001, xingIterativeMethodsInverse1996, yuIntensitymodulatedArcTherapy2011}. This optimization problem is typically defined as a constrained optimization minimizing a weighted cost function that balances multiple conflicting objectives \cite{webbPhysicalBasisIMRT2003, xingOptimizationImportanceFactors1999,yangClinicalKnowledgebasedInverse2004}. The result is greatly impacted by the value of the optimization parameters (e.g., the objective weights in the cost function), often resulting in plans that cannot meet all objectives at once. 

A main challenge in treatment planning is translating the overall clinical goals into weighted objective functions and dose constraints yielding acceptable plans. Human planners resort to a trial-and-error approach, whereby they iteratively adjust the optimization parameters based on the results from the optimization problem until the resulting plans meet the clinical requirements \cite{xingOptimizationImportanceFactors1999}. Treatment planning is potentially subjective, since changes in the treatment parameters are based on the experience of the planner \cite{husseinAutomationIntensityModulated2018a}. Additionally, trial-and-error involves repeatedly using the computationally-expensive optimization algorithms over many iterations, with potentially multiple interactions between planner and radiation oncologists in between, making the entire process time-consuming and costly, and presenting a bottleneck in precision RT.

Multiple approaches such as knowledge-based planning or multi-criteria optimization have been investigated to reduce the load on human planners \cite{husseinAutomationIntensityModulated2018a}. Knowledge-based approaches utilize prior experience to improve plans or derive good starting points for subsequent trial-and-error adjustment. Some examples include selecting closest matching patients \cite{petrovicKnowledgelightAdaptationApproaches2016, mcintoshContextualAtlasRegression2016, yangClinicalKnowledgebasedInverse2004,huangLearningImageRepresentations2023}, or using models that predict achievable dose volume histograms (DVHs) \cite{munterDosevolumeHistogramPrediction2015}. Alternatively, the multi-criteria approach often involves a ``wish list'' containing objective functions with assigned priorities, which iteratively steers planning to meet as many objectives as possible ordered by their priority \cite{breedveldICycleIntegratedMulticriterial2012a, breedveldNovelApproachMulticriteria2007, voetFullyAutomatedMulticriterial2013, voetIntegratedMulticriterialOptimization2012}. 

With the recent success of deep learning, multiple neural network-based approaches have also been proposed to assist treatment planning, mostly based on U-net \cite{ronnebergerUNetConvolutionalNetworks2015} or generative adversarial frameworks \cite{goodfellowGenerativeAdversarialNets2014}. Most works focus on predicting dose distributions that may be achievable based on previous clinical plans. As an input, these models take organ masks \cite{maDoseDistributionPrediction2019, fanAutomaticTreatmentPlanning2019, nguyenFeasibilityStudyPredicting2019, maIncorporatingDosimetricFeatures2019}, possibly also including beam information \cite{barragan-monteroThreeDimensionalDose2019}. Furthermore, these learning-based methods require large datasets and extensive training to ensure generalization across disease sites and device settings, making it difficult to integrate them into clinical workflow.

Adjusting the optimization parameters can be defined as a decision-making problem that can ideally be fully automated. Framing treatment planning as a sequence of actions taken by a planner, reinforcement learning (RL) has been recently applied to mimic human planners in RT workflows \cite{xuDeepReinforcementLearning2022}. RL methods learn in a trial-and-error manner how to take actions maximizing a cumulative reward (i.e., a scalar based on dose metrics or target conformity). Some initial studies demonstrate that RL agents can learn to modify optimization parameters such as weights and objective doses \cite{shenOperatingTreatmentPlanning2020}, while other works train agents that directly optimize machine parameters \cite{hrinivichArtificialIntelligencebasedRadiotherapy2020}. In practice, RL-based approaches face multiple problems. First, designing reward functions of RL that balance between different
objectives — also known as the ``reward design problem'' - is notoriously difficult because agents are susceptible to reward hacking\rebutal{, which is when they exploit flaws or loopholes in the reward function to achieve high rewards in unintended ways, rather than learning the desired behavior.}~\cite{norvig2002modern, wiering2012reinforcement}. Indeed, it is not straightforward how to write a reward function to evaluate the treatment plan that requires balancing multiple clinical objectives. While RL agents can
learn from labeled examples, this is not possible with a single example; RL agents need to gather
large amounts of labeled data to capture the nuances of different users’ preferences and objectives, which
has shown to be costly~\cite{zhang2016learning}.  Additionally, both \rebutal{learning and RL based methods}  do not generalize well to new users or new environments that have different objectives, necessitating re-design or re-collection of labeled data.

Our aim is to leverage multi-modal \rebutal{large language models (LLMs)} to create an easier way for RT professionals to communicate their preferences and to facilitate the inverse treatment planning process. In the past few years, LLMs that are trained on internet-scale data have witnessed significant advancements and shown an impressive ability to learn in-context from a few examples~\cite{liu2023context}. Our key insight is that the scale of data that multi-modal LLMs have been trained on makes them great in-context learners~\cite{liu2023context}, also allowing them to capture meaningful commonsense priors about human behavior as well as expertise on specific domains. Given a few examples and/or a description demonstrating the radiation oncologist’s objective, an LLM should be able to provide an accurate evaluation of a new plan and provide solutions for further improvement.

In this work, we explore how to prompt multi-modal LLMs as a proxy reward function (evaluator) and agent (planner) for radiotherapy treatment planning. Specifically, we propose a novel framework to fully \rebutal{automate the planning process} using pre-trained multi-modal LLMs. The proposed GPT-RadPlan is capable of learning to mimic human planners without any domain-specific training, using only three clinical plans as a reference, in the form of \rebutal{optimized weights, DVH tables (the DVH in a tabular format where each row contains doses received for a given fraction of the volume of each of the organs)}, and dose distribution images. GPT-RadPlan iteratively adjusts optimization parameters, directly assessing whether the plan meets the clinical requirements based on agreement with the clinical protocol. In this case, GPT-RadPlan acts both as an evaluator and planner, first determining which aspects of the plan can be improved on the clinical requirements, and subsequently suggesting new optimization weights to address the identified deficiencies. Through our experiments, we compare GPT-RadPlan's results to clinically approved \rebutal{treatment plans for prostate and head \& neck cancer (HN) patients}.  There are two advantages to prompting multi-modal LLMs for treatment planning: (1) we can leverage LLM’s in-context learning abilities and prior knowledge of treatment planning so that users only need to provide a handful of examples of desirable plans, and (2) we can specify our preferences intuitively using \rebutal{human language in text}. 

\begin{figure*}[t!]
    \centering
    \includegraphics[width=\textwidth]{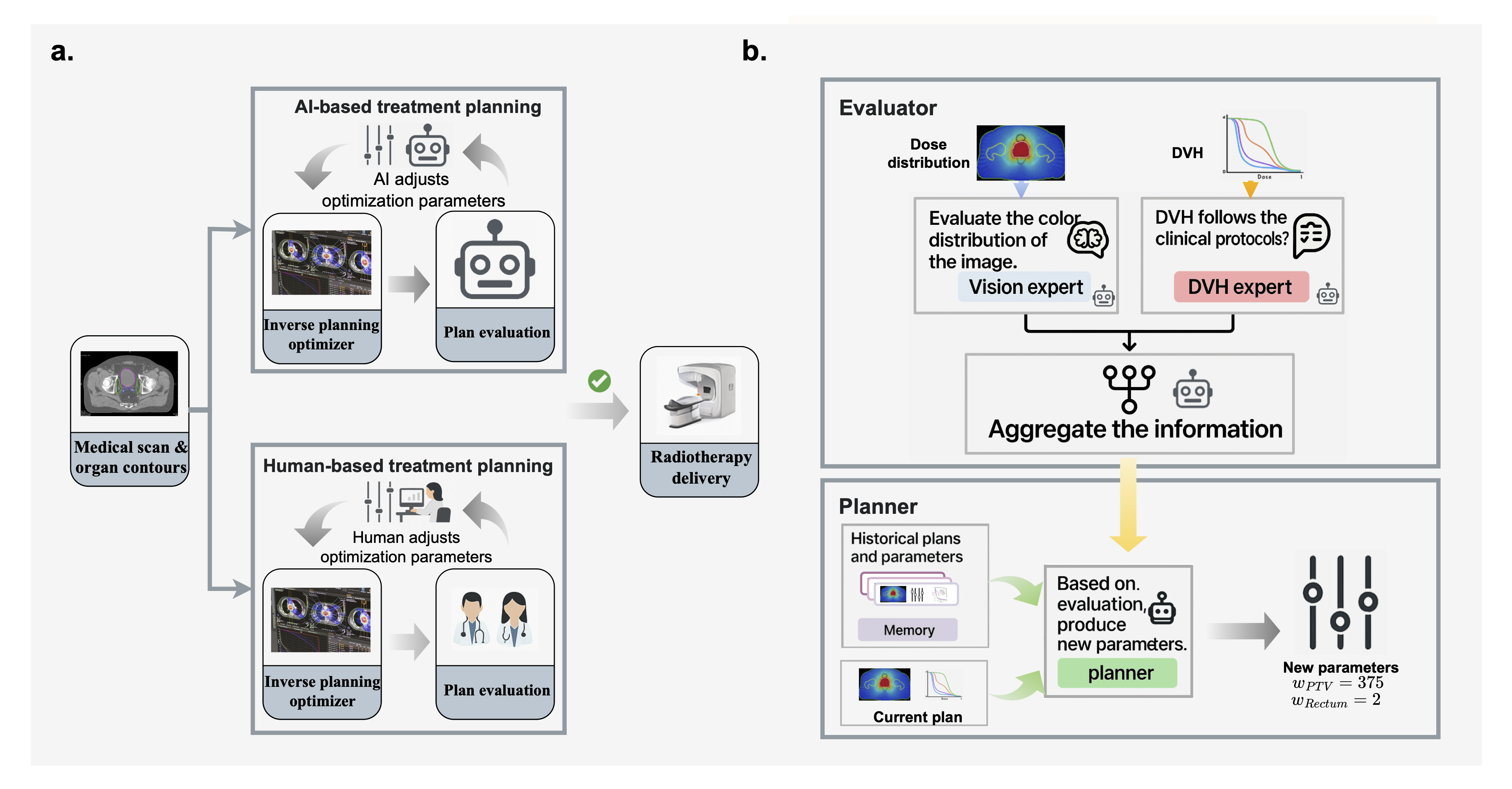}
    \caption{\textbf{Overview of GPT-RadPlan workflow.} a. Integration of multi-modal LLM-based RT treatment planning into the existing clinical workflow, using GPT-4V. GPT-RadPlan can \rebutal{be human-in-the-loop where the users can use human language to provide feedback on how to improve the current plan when needed,} and iteratively adjust optimization weights. b. Modules for the application of GPT-4V at different steps of the RT workflow. Based on the independent evaluation of the dose distribution by an image AI expert, and the DVH assessment from the DVH AI expert, GPT-4V reviews the plan, providing feedback to the AI planner, \rebutal{and approves the plan after the plan is good enough}.  The planning module translates feedback from the evaluation module into concrete optimization parameters, enabling iterative refinement of the treatment plan. GPT-4V's in-context learning capabilities facilitate this process without requiring additional model training, relying instead on input prompts and historical planning data.}
    \label{fig:arch}
\end{figure*}

\section{Methodology}

\subsection{Problem formulation}

We opt for a two-loop optimization formulation \cite{xingOptimizationImportanceFactors1999, huangMetaoptimizationFullyAutomated2022} consisting of (i) \rebutal{an inner loop performing inverse fluence optimization} and (ii) an outer loop step optimizing the parameters of the inner loop. As such, the inner loop performs traditional fluence map optimization, \rebutal{trying to obtain the optimal fluence map} $x$ by minimizing a cost function composed of multiple weighted objectives for different targets and organs at risk, defined as

\begin{equation}
    \begin{aligned}
        \min_{x} \quad & \sum_p w_{p}([Ax]_p - d_{\rebutal{t}})^2 + \sum_s w_{s}\Theta([Ax]_s - d_{s})([Ax]_s - d_{s}),\\
        \textrm{s.t.} \quad &D_{95}([Ax]_p)=d_p\\
          & x\geq0.     \\
    \end{aligned}
    \label{eq:inner}
\end{equation}

Here, $\{w_p\}_{p=1}^{N_p}$ and $\{w_s\}_{s=1}^{N_s}$ are the weights that balance the different objectives for $N_p$ PTV targets and $N_s$ \rebutal{organs at risk (OARs)}, respectively. $A$ represents the dose influence matrix, containing the dose per unit fluence delivered to voxels in the volume by beamlets, and $\{d_p\}_{p=1}^{N_p}$ and $\{d_s\}_{s=1}^{N_s}$ are the scalar objective doses for each structure. Essentially, the whole cost function penalizes squared deviations from the target objective doses, regardless of the positive or negative sign, while the OAR terms only penalize squared over-dosing for doses above $d_s$. As such, $\Theta$ represents the Heaviside function constraining the objective to only positive values. The minimization is constrained for only positive $x$ values, and forcing that the $D_{95}$ --- \rebutal{the minimum dose that covers at least 95\% og the target volume} --- equals the prescribed dose for the targets.

We design GPT-RadPlan to operate the outer loop, finding the optimal target and OAR weights that are used to minimize Equation \ref{eq:inner}. GPT-RadPlan consists of three algorithmic components: \rebutal{a) an evaluation module that leverages GPT-4V prior knowledge as well as reference plans to evaluate the plan in the current iteration, b) a memory module that stores the plans and the corresponding optimization parameters in the previous iterations, c) a planning module that converts the text-based feedback from the evaluation module to PTV and OAR weights for inverse planning optimizer.}

\subsection{Overall framework}

As shown in Figure~\ref{fig:arch}, GPT-RadPlan introduces LLM-based agents as the evaluator and planner. The system incorporates three fundamental modules to facilitate the treatment planning process, enabling iterative reasoning, planning, and continuous interaction between the agents and the treatment planning system. The task concludes either when the patient-specific requirements in the physician's intent are successfully achieved or when the maximum iteration limit is reached. Algorithm~\ref{alg:gpt_radplan} illustrates the detailed algorithm which breakdowns the GPT-RadPlan step-by-step. The functionalities of each module are as follows:

\begin{algorithm}[t]
    \caption{GPT-RadPlan Auto-Plan Framework}
    \label{alg:gpt_radplan}
    
    \textbf{Input:} Patient data ($P$), Reference plans ($R$), Clinical requirements ($C$), Max iterations ($N$) \\
    \textbf{Output:} Optimized parameters $w$
    
    \begin{algorithmic}[1]
    \State Initialize \textbf{EvaluationModule} with reference cases and clinical objectives
    \State Initialize \textbf{MemoryModule} for storing historical trajectories
    \State Initialize \textbf{PlanningModule} for generating optimization parameters
    \State Extract initial dose distribution $I_0$ and DVH table $\delta_0$ from patient data
    
    \For{$i = 1$ to $N$}
        \State Feedback $F_i \gets \textbf{EvaluationModule}.\text{evaluate}(I_{i-1}, \delta_{i-1})$ \Comment{Evaluate current plan}
        \State \textbf{MemoryModule}.\text{store}($I_{i-1}, \delta_{i-1}, F_i$) \Comment{Store the current state in memory}
        \State $w \gets \textbf{PlanningModule}.\text{plan}(F_i, \textbf{MemoryModule}.\text{getRecentTrajectory}())$
        \Comment{Adjust parameters based on feedback and memory}
        \State $I_i, \delta_i \gets \text{innerLoopOptimization}(w)$ \Comment{Generate new plan using updated parameters}
        \If{$\textbf{EvaluationModule}.\text{isPlanSatisfactory}(I_i, \delta_i)$}
            \State \textbf{break} \Comment{Stop if the plan meets clinical objectives}
        \EndIf
    \EndFor
    \State \Return $w$
    \end{algorithmic}
\end{algorithm}

\paragraph{Evaluation module}
The evaluator assumes a central role within the workflow of GPT-RadPlan, assessing the level of agreement between the current plan and the physician's intent. It receives the present dose distribution images, DVHs, and dose statistics \rebutal{such as the precomputed minimum, maximum, and mean doses based on DVHs} together with the physician's intent (in textual form as a list of requirements). Functioning as a centralized coordinator, the evaluator engages in reasoning and providing suggestions on directions of improvement for the current plant. Such suggestions from the \rebutal{evaluation module} are then used to adjust the optimization parameters and then optimize the plan, allowing for a continuous cycle of plan improvement. 

To properly evaluate both qualitative aspects about the dose distribution and quantitative values and DVHs of the treatment plan, we separately employ GPT-4V as a vision and DVH expert. The vision expert interprets the dose distribution images, analyzing the regional color intensity \rebutal{based on the color bar}, presence of hot or cold spots, and distribution and spillage around the PTV and OARs. This analysis is done using dose images were \rebutal{automatically cropped around each structure, using the contours with a 5-pixel margin}, as well as the full dose distribution in the sagittal, coronal, and axial axes.

On the other hand, the DVH expert will take the DVH and related dose statistics as input. The DVH is first transformed into a table and then translated into a text sequence. The corresponding physician's intent with prescribed dose and clinical requirements for the patient is also provided. The model then reasons about the level of agreement of the DVH and metrics with the intent. Aggregating the outputs of the vision expert and DVH expert, the GPT-4V provides final feedback on how to improve the current treatment plan.

For a treatment plan, we obtain its dose distribution images $I$ and the DVH $\delta$, the evaluator then evaluates them to derive feedback $F$ based on the prompt as $F = \text{EvaluationModule}(I,\delta).$

\paragraph{Memory module} The memory module serves to store crucial information needed during the planning process to aid the accumulation of useful knowledge and enhance the agent’s decision-making capabilities. Specifically, the memory module is used to store the \rebutal{plans and the corresponding optimization parameters from past iterations for the same patient}. The memory module also incorporates a mechanism for filtering redundant information. If the planning process \rebutal{takes too many iterations}, it retains only the most recent $L$ steps of dose distribution images, DVHs, feedback and the optimization parameters, forming a set in the order of steps $\{(I_{t-L+1}, \delta_{t-L+1}, F_{t-L+1}, w_{t-L+1}),\dots, (I_{t}, \delta_{t}, F_{t}, w_{t})\}$. This assists the agent in comprehending the relationship between the optimization parameters and changes in the treatment plan.

\paragraph{Planning module} 
The planning module fulfills the vital function of converting the text-based feedback from the evaluation module to optimization parameters that can be processed by the treatment planning system. The planning module receives the current iteration of the treatment plan and the feedback from the evaluation module and extracts pertinent details from the memory module, enabling learning from its historical trajectories.

To help with initialization and guide the parameter adjustment process, the planning module receives some reference examples of previously approved treatment plans \rebutal{from the same treatment site and clinical protocol}, together with the optimization parameters used to obtain these plans.  This further facilitates the planning module's understanding of the relationship between the optimization parameters and changes in the treatment plan. As a result, the model outputs the optimization parameters that will drive the inner optimization loop of the next iteration $w = \text{PlanningModule}(I_t,\delta_t,F_t)$. \rebutal{The example prompts can be found in~Section \ref{sec:tp_prompts} of the supplementary.}

\subsection{Dataset}
This study utilizes a dataset comprising imaging and treatment plans for 17 prostate cancer patients and 13 head \& neck cancer patients who underwent VMAT. Available data for each patient includes CT scans, delineated anatomical structures, and clinically approved treatment plans \rebutal{created} via Eclipse\textsuperscript{\tiny\textregistered} (\rebutal{version 15.6}, Varian Medical Systems, Palo Alto, CA).

For \rebutal{the} prostate cancer patients, all cases adhere to a clinical protocol designed to deliver \rebutal{the mean dose of 70.2 Gy} to the PTV. Concurrently, the protocol stipulates specific dose constraints for the OARs, which include the bladder, rectum, and both femoral heads. These dose limits typically involve \rebutal{constraints} on the mean OAR doses and DVH constraints, such as restricting no more than 34\% of the rectum to receive in excess of 31 Gy.

Treatment plans for \rebutal{the} head \& neck cancer patients follow a protocol that prescribes a dose of 70 Gy to the \rebutal{primary PTV (PTV70). Additionally, two nodal PTVs — PTV56 and PTV52 — receive prescribed doses of 56 Gy and 52 Gy, respectively. All these doses are prescribed to at least 95\% of the target volume.} Key OARs for this group are the brainstem, spinal cord, oral cavity, larynx, and both parotid glands. The dose constraints for these OARs include upper limits on the mean dose received, with several organs also subjected to maximum dose limits.

\subsection{Technical details and evaluation}
We develop an in-house two-loop optimization system, where GPT-RadPlan performs outer loop optimization of the optimizer weights, and the inner loop fluence map optimization is based on the open-source \rebutal{treatment planning system (TPS)} matRad \cite{wieserDevelopmentOpensourceDose2017}. All plans utilize a CT, \rebutal{and the dose grid had a voxel size} of $2.5\times2.5\times2.5 \text{mm}^3$ resolution. MatRad VMAT plans consist of \rebutal{1 to 3 full} co-planar arcs and are based on the SmartArc planning algorithm \cite{macfarlaneTechnicalNoteFast2020, christiansenContinuousApertureDose2018}, using the interior-point optimization (IPOPT) package for the inverse optimization step.

Throughout our experiments, we compare the plans generated by GPT-RadPlan to \rebutal{plans optimized by multiple baseline methods and human-designed clinically approved plans}. These include the plans generated by certified physicists from our institution with multiple years of experience in developing treatment plans across various cancer types, including prostate and head \& neck cases. We also include plans obtained as a result of running an automated planning algorithm~\cite{yangEvaluationKnowledgeGuidedAutomated2020}, referred to as Autoplan in the remainder of the manuscript, which is a knowledge-guided iterative algorithm based on a dose-anatomy guided voxelization scheme. \rebutal{The Autoplan script first sets up all the treatment planning parameters automatically based on the prescription and disease sites. The prediction model based on the dosimetric metric from the simple treatment plans (PTV-only plans) is then used to predict and initialize the personalized optimization constraints automatically for a new patient with the script. The script includes a plan evaluator for quantitative plan assessment and the algorithm to automatically update treatment plan parameters in each iteration. These perform experience-based adjustments, first checking the dose/volume with the preset constraints and subsequently modifying the weights and dose constraints based on difference between the results of the current iteration and the clinical requirements \cite{yangEvaluationKnowledgeGuidedAutomated2020}.}

The \rebutal{third} baseline, referred to as Bayesian Optimization (BO), where we adopted the plan quality metrics (PQM) mentioned in~\cite{wang2023high} as the objective in the outer loop and used Optuna~\cite{akiba2019optuna}\rebutal{, a commonly used package designed for Bayesian optimization}, to implement Bayesian optimization for optimizing parameters. Since the original paper only presented this BO method and provided the model configuration for prostate cases, we only apply BO to our prostate cohort.

To compare plans, we use multiple dose metrics and indexes that are frequently used in the clinic. These include the \rebutal{the mean dose}, $D_{95}$, homogeneity index (HI) and conformity index (CI) and for the targets, and the $D_{5}$, $D_{50}$, $V_{15}$ and $V_{30}$ for the OARs. 

Here, we provide a brief description of each of these metrics. First, we utilize the $D_{q}$ --- where $q$ is a given percentage from 0\% to 100\% --- to denote the common maximum dose received by $q\%$ of the target/organ volume. Similarly, the $V_{d}$ represents the percentage of the volume receiving at least a dose of $d$ Gy. For targets, both a higher $D_{q}$ close to the prescribed dose and a higher $V_{d}$ close to 100\% are desirable. Conversely, for OARs, both the $D_{q}$ and $V_{d}$ metrics would ideally be as low as possible.

Specifically for the target, the HI indicates whether the PTV is uniformly covered with a dose close to the prescribed value. In this study, we calculate the HI based on the $D_5$ and $D_{95}$ values as

\begin{equation}
    HI = \frac{D_5 - D_{95}}{d_p}\times 100,
\end{equation}

\noindent where lower values close to 0 indicate ideal dose homogeneity. The CI measures both the conformation of doses on target and the volume of surrounding tissue that is covered by the reference dose. While multiple methods have been proposed to calculate this CI, we opt for the following definition

\begin{equation}
    CI = \frac{(TV_{95,PTV})^2}{TV\times TV_{95,Body}},
\end{equation} 

\noindent where $TV_{95,PTV}$ and $TV_{95,Body}$ denotes the total volumes receiving at least 95\% of the prescribed dose for the PTV and the whole body, respectively, and $TV$ is the total \rebutal{target} volume. In the ideal scenario, the volume receiving the prescribed dose is confined to just the PTV, and the CI tends to 1 as the numerator and denominator become equal.

Based on these metrics, we perform multiple experiments to (i) determine the optimal architecture for GPT-RadPlan's evaluation module, and (ii) assess the quality of GPT-RadPlan's generated plans, comparing them to the available clinical plans.

\section{Results}

\subsection{Planning performance}
To assess GPT-RadPlan capabilities in mimicking clinical human planners, we first perform a quantitative comparison between GPT-RadPlan generated plans and clinical plans based on the aforementioned metrics. We also compare the types of plans based on the final DVH and dose distributions.

Table \ref{tab:ptv} contains the average PTV metrics across all plans for both disease sites. For the prostate cases, GPT-RadPlan outperforms all baselines including clinical human planners across all metrics, achieving a higher minimum dose, a $D_{95}$ that exactly matches the prescribed dose, and \rebutal{significantly lower compared} dose homogeneity and conformity. In the head \& neck cases, GPT-RadPlan matches or outperforms the baselines, showing higher minimum doses and better conformity. Additionally, Table \ref{tab:oar} displays the organ-sparing metrics for the main OARs. For both the prostate and head \& neck cases, \rebutal{GPT-RadPlan produces treatment plans that generally deliver lower doses to OARs compared to the baselines, as reflected by improved values in most evaluation metrics.}

\begin{table}[]
\caption{\textbf{PTV dose metrics}. Several dose metrics of the PTV target are displayed for all the GPT-RadPlan plans and baselines, including the mean and minimum doses, as well as the $D_{95}$, the HI and the CI. For all the metrics, we include the average values across plans and the standard deviation in brackets. Values in \textbf{bold} represent the \rebutal{statistically best result for each PTV target, when the average minus the standard deviation is higher than the average plus the standard deviation of the second best value.}}
\footnotesize
    \begin{tabular}{@{}lllllll@{}}
    \toprule
    \textbf{Target} & \textbf{Method}   & \textbf{Mean dose [Gy]} & \textbf{Min dose [Gy]}  & \textbf{$\textbf{D}_{95}$ [Gy]} & \textbf{HI} & \textbf{CI}   \\ \midrule
    
    \multicolumn{7}{c}{Prostate plans} \\\midrule
    \multirow{4}{*}{PTV70.2} 
                             & Clinical & 72.47 (0.64) & 61.34 (2.21) &  70.04 (0.18) & 5.43 (1.04) & 0.883 (0.020) \\
                             & Autoplan & 72.78 (0.52) & 61.83 (1.24) &  70.03 (0.28) & 6.01 (0.60) & 0.870 (0.011) \\
                             & BO & 73.04 (1.65) & 60.66 (2.28) &  69.76 (0.46) & 5.05 (0.83) & 0.875 (0.016) \\
                             & GPT-RadPlan & \textbf{70.94} (0.65) & \textbf{63.82} (2.19) &  \textbf{70.21} (0.06) & \textbf{2.43} (1.11) & \textbf{0.912} (0.041) \\ \midrule
                             
    \multicolumn{7}{c}{Head \& neck plans} \\\midrule
    \multirow{3}{*}{PTV70} 
                           & Clinical & 71.85 (0.26) & 60.57 (3.72) & 69.69 (0.14) & 5.05 (0.58) & 0.881 (0.048) \\
                           & Autoplan & 72.05 (0.27) & 62.12 (2.73) & 69.82 (0.17) & \textbf{5.01} (0.47) & 0.871 (0.038) \\
                           & GPT-RadPlan & \textbf{72.73} (0.30) & \textbf{62.73} (1.10)  & \textbf{70.00} (0.01) & 7.12 (0.71) & \textbf{0.948} (0.005) \\ \midrule
                           
    \multirow{3}{*}{PTV56} & Clinical & 65.47 (3.53) & 41.40 (9.71)  & 56.43 (2.09) & - & - \\
                           & Autoplan & 67.58 (2.71) & 49.07 (10.9)  & 57.45 (0.60) & - & - \\
                           & GPT-RadPlan & 64.66 (4.51) & \textbf{49.62} (2.18)  & 57.31 (1.39) - & - \\ \midrule
                           
    \multirow{3}{*}{PTV52} & Clinical & 54.58 (0.80) & 44.57 (2.01)  & 52.14 (0.53) & - & - \\
                           & Autoplan & 55.44 (0.44) & 44.08 (1.73)  & 51.89 (0.46) & - & - \\
                           & GPT-RadPlan & \textbf{54.26} (0.45) & \textbf{45.04} (3.09) & 52.42 (0.24) & - & - \\ \bottomrule
    \end{tabular}
    \centering
    \label{tab:ptv}
\end{table}

\begin{table}[]
\caption{\textbf{OAR dose metrics}. Multiple dose metrics capturing OAR sparing in prostate and head \& neck plans are presented. These include the mean dose, the $D_5$ and $D_{50}$ (maximum common dose that 5\% or 50\% of the volume receives), and the $V_{15}$ and $V_{30}$ (representing which percentage of the volume receives 15 and 30 Gy, respectively). For all the metrics, we include the average values across plans and the standard deviation in brackets. In all cases, lower values are preferred.}
\footnotesize
    \begin{tabular}{@{}lllllll@{}}
    \toprule
    \textbf{Organ} & \textbf{Method} & \textbf{Mean dose [Gy]} & \textbf{$\textbf{D}_{5}$ [Gy]} & \textbf{$\textbf{D}_{50}$ [Gy]} & \textbf{$\textbf{V}_{15}$ [\%]} & \textbf{$\textbf{V}_{30}$ [\%]} \\ \midrule
    
    \multicolumn{7}{c}{Prostate plans} \\\midrule
    \multirow{4}{*}{Bladder} & Clinical & 24.48 (9.30) & 62.14 (10.79) & 18.87 (12.12) & 60.85 (29.52) & 34.45 (15.28) \\
                             & Autoplan & 31.91 (1.11) & 59.48 (7.71) & 29.29 (0.87) & 94.67 (3.92) & 47.08 (2.72) \\
                             & BO & 23.88 (6.45) & 64.26 (10.00) & 20.04 (5.50) & 54.13 (21.17) & 35.18 (12.11)\\
                             & GPT-RadPlan & 17.02 (8.11) & 60.08 (14.39) & 9.18 (8.63) & 35.60 (17.07) & 22.62 (13.12) \\ \midrule
                             
    \multirow{4}{*}{Rectum} & Clinical & 26.45 (5.64) & 63.64 (8.82) & 22.20 (5.71) & 71.91 (14.92) & 32.87 (10.54) \\
                            & Autoplan & 25.73 (3.19) & 57.63 (8.76) & 21.13 (2.26) & 78.35 (10.45) & 27.07 (6.66) \\
                            & BO & 22.39 (5.55) & 67.81 (6.44) & 22.13 (8.42) & 75.32 (13.17) & 28.14 (9.13)\\
                            & GPT-RadPlan & 21.80 (6.60) & 58.88 (13.81) & 17.09 (7.99) & 50.44 (20.03) & 27.65 (16.43) \\ \midrule

    \multicolumn{7}{c}{Head \& neck plans} \\\midrule
    \multirow{3}{*}{Brainstem} & Clinical & 12.17 (4.29) & 30.95 (4.78) & 8.34 (5.75) & 33.36 (14.10) & 10.03 (11.03) \\
                               & Autoplan & 9.85 (3.41) & 27.47 (5.84) & 6.24 (3.42) & 25.22 (9.81) & 5.67 (5.86) \\
                               & GPT-RadPlan & 10.95 (4.89) & 26.91 (4.74) & 8.09 (7.51) & 32.98 (16.07) & 0.03 (0.11) \\ \midrule
                               
    \multirow{3}{*}{Larynx} & Clinical & 42.80 (11.03) & 69.95 (4.30)	 & 39.85 (14.87) & 97.44 (3.24) & 64.06 (18.18) \\
                            & Autoplan & 41.89 (9.95) & 69.08 (4.79) & 37.13 (13.30) & 99.94 (0.17) & 60.92 (15.95) \\
                            & GPT-RadPlan & 34.15 (14.33) & 70.68 (5.75) & 26.96 (21.64) & 64.12 (26.95) & 42.46 (25.01) \\ \midrule
                            
    \multirow{3}{*}{Oral Cavity} & Clinical & 34.86 (4.15) & 68.07 (8.83) & 33.62 (12.29) & 92.96 (5.28) & 50.00 (11.01) \\
                                & Autoplan & 35.61 (4.92) & 68.08 (9.77) & 30.96 (4.59) & 97.39 (3.65) & 51.06 (13.87) \\
                                 & GPT-RadPlan & 25.41 (8.51) & 66.38 (15.18) & 18.24 (8.33) & 54.32 (25.66) & 27.89 (13.78) \\ \midrule
                                 
    \multirow{3}{*}{Parotid Left} & Clinical & 35.10 (13.93) & 63.18 (10.85) & 33.50 (21.21) & 71.64 (20.97) & 52.80 (24.12) \\
                                & Autoplan & 31.70 (11.77) & 63.34 (10.52) & 30.12 (20.03) & 61.94 (16.5) & 44.58 (19.62) \\
                                  & GPT-RadPlan & 24.26 (12.65) & 57.76 (16.06) & 19.73 (18.61) & 42.99 (20.67) & 28.49 (21.04) \\ \midrule
                                  
    \multirow{3}{*}{Parotid Right} & Clinical & 34.32 (10.67) & 61.69 (11.30) & 32.22 (13.83) & 79.00 (21.05) & 52.65 (20.65) \\
                                    & Autoplan & 29.93 (9.81) & 61.17 (12.90) & 27.51 (14.09) & 62.63 (13.03) & 41.85 (17.51) \\
                                   & GPT-RadPlan & 21.66 (7.79) & 58.49 (15.56) & 13.66 (7.45) & 41.84 (19.71) & 25.60 (13.81) \\ \midrule
                                   
    \multirow{3}{*}{Spinal Cord} & Clinical & 17.36 (6.32) & 33.88 (5.15) & 18.84 (13.24) & 55.32 (13.69) & 24.83 (25.69) \\
                                & Autoplan & 14.91 (3.04) & 32.07 (1.62) & 13.74 (9.21) & 53.00 (14.41) & 16.00 (7.62) \\
                                 & GPT-RadPlan & 16.27 (3.96) & 30.14 (0.24) & 17.76 (12.52) & 55.78 (14.19) & 1.34 (3.25) \\ \bottomrule
    \end{tabular}
    \centering
    \label{tab:oar}
\end{table}

\begin{figure}[t!]
    \centering
    \begin{subfigure}[]{\textwidth}
        \centering
        \includegraphics[width=\textwidth]{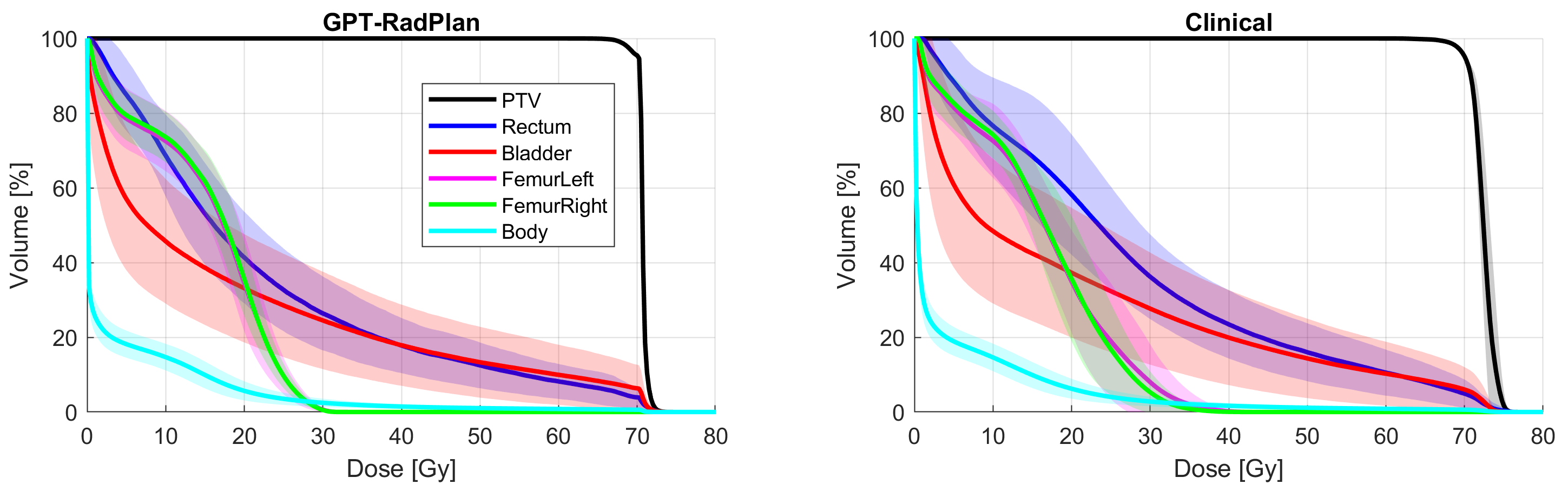}
        \caption{Prostate DVHs}
        \label{fig:dvh_prostate}
    \end{subfigure}
    \begin{subfigure}[]{\textwidth}
          \centering
          \includegraphics[width=\textwidth]{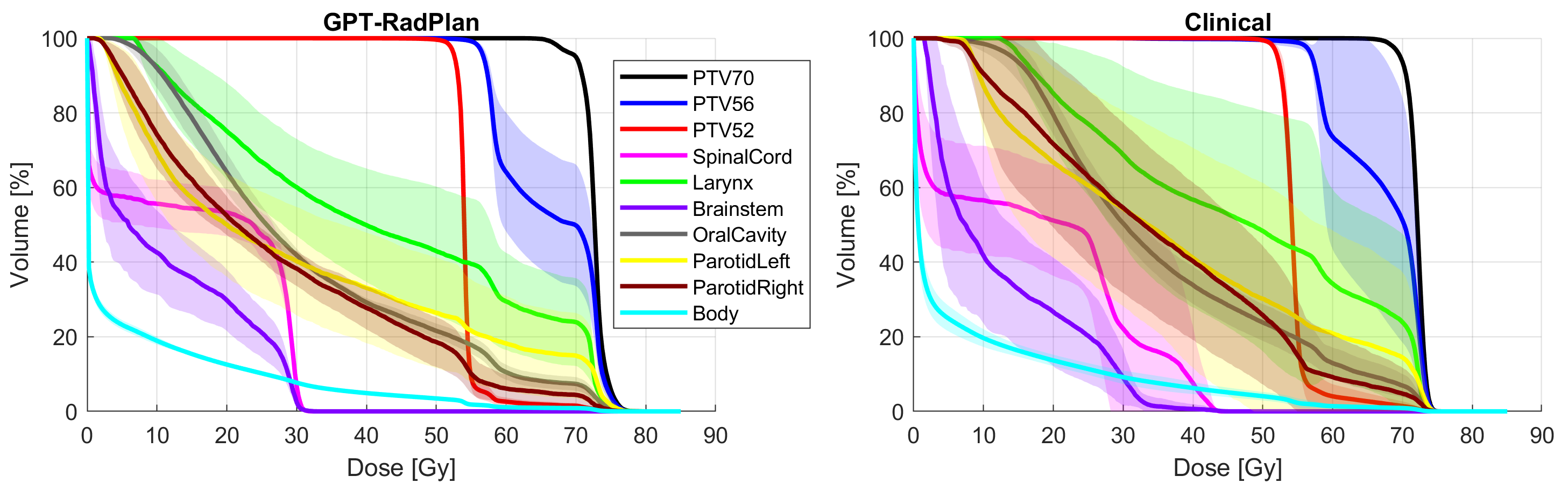}
          \caption{Head \& neck DVHs}
          \label{fig:dvh_hn}
    \end{subfigure}
    \caption{\textbf{DVH comparison}. Visual comparison of the DVHs for (left) GPT-RadPlan plans, and (right) clinical plans. Solid lines show the mean values, while the shaded bands indicate the standard deviation. For the OARs, better plans are usually characterized by lines that are close to the bottom left corner, implying greater OAR sparing.}
    \label{fig:dvh}
\end{figure}

\begin{figure}[]
    \centering
    \includegraphics[width=\textwidth]{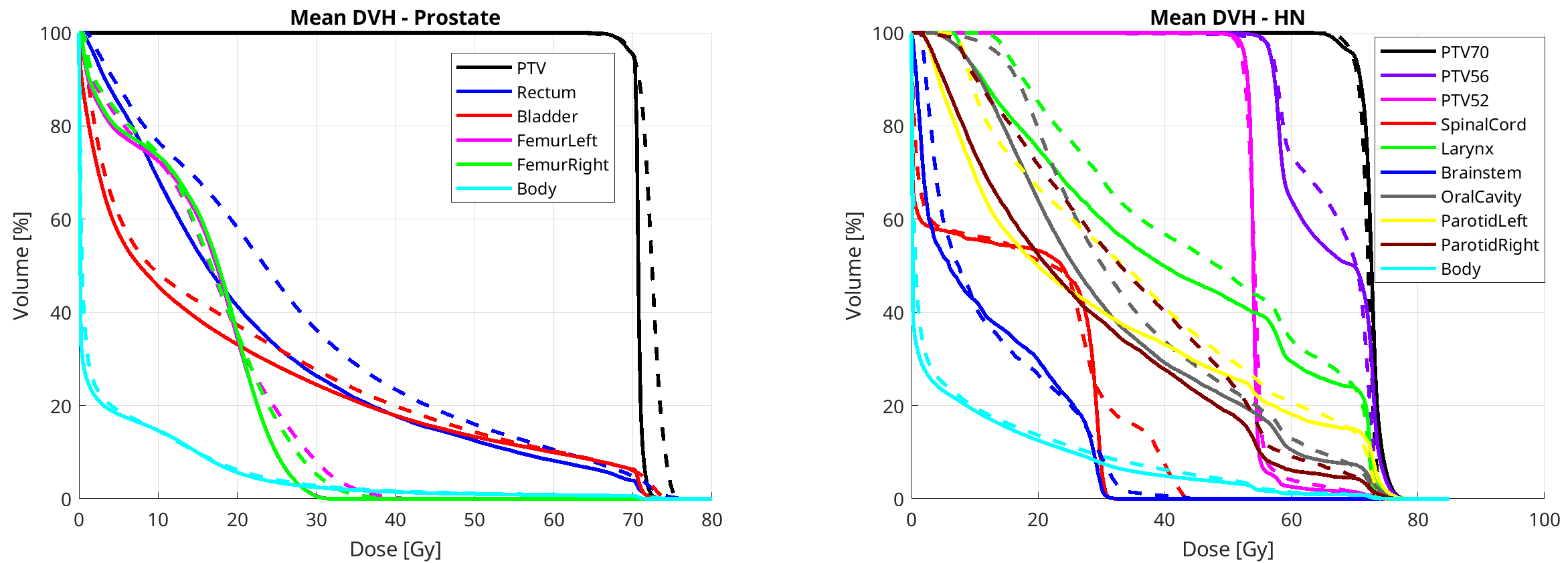}
    \caption{\rebutal{\textbf{Average DVH comparison}. Visual comparison of the average DVH lines across all prostate (left) and lung (right) patients. Solid lines represent GPT-RadPlan plans, while dashed lines indicate clinical plans.}}
    \label{fig:avg_dvh}
\end{figure}

Figures \ref{fig:dvh} \rebutal{and \ref{fig:avg_dvh}} further confirms GPT-RadPlan’s superiority in sparing OARs. In the prostate cases depicted in Figure \ref{fig:dvh_prostate}, GPT-RadPlan achieves a sharper dose fall-off within the PTV (black line), resulting in more homogeneous delivery. Additionally, GPT-RadPlan generates plans that considerably spare the rectum compared to their clinical counterparts, while maintaining comparable doses to the femoral heads and slightly decreasing average bladder doses. In the head \& neck cases illustrated in Figure \ref{fig:dvh_hn}, the clinical plans generally result in lower doses to all OARs, particularly for the larynx, oral cavity, and parotid glands, which often partially overlap with the target volumes. While the shaded areas in the clinical plans may indicate \rebutal{dose} spillage in some head \& neck plans, the physician’s intent only specifies a lower dose limit for PTV56 and PTV52, with no upper limit. While the human planners primarily focused on enforcing this lower limit, GPT-RadPlan employs an optimization objective that penalizes both positive and negative deviations from the 56 Gy target, which likely explains the differences observed in the DVH distribution.

To statistically demonstrate the superiority of our method over clinical plans in reducing \rebutal{OAR} mean doses, we performed a Wilcoxon signed-rank test, a non-parametric test for paired samples \rebutal{which is commonly used to prove whether paired measurements are different without assuming normal distribution of the data}. This test evaluates whether the \rebutal{median of the mean doses extracted from the different plans obtained from GPT-RadPlan} are significantly lower than those of \rebutal{the} clinical plans. Specifically, a one-tailed test was conducted with the null hypothesis stating that GPT-RadPlan method does not achieve a lower mean dose compared to clinical plans. The test provides p-values, and the null hypothesis is rejected when the p-value is less than the 0.05 significance level, which indicates that GPT-RadPlan achieves statistically significant improvements in mean dose reduction. After running the Wilcoxon test, we obtain a p-value of 0.003 and 0.001 for the rectum and bladder, demonstrating that GPT-RadPlan delivers less dose to these organs. Similar results are obtained for the head \& neck cases, with p-values of 0.06, 0.006, 0.0001, 0.0001, 0.0003 and 0.729 for the brainstem, larynx, oral cavity, left and right parotid and spinal cord, respectively. These values further confirm that GPT-RadPlan can reduce the dose to most OARs except for the brainstem and spinal cord.

\begin{figure*}[t!]
    \centering
    \begin{subfigure}[]{\textwidth}
        \centering
        \caption{\textbf{Prostate patients}}
        \includegraphics[width=\textwidth]{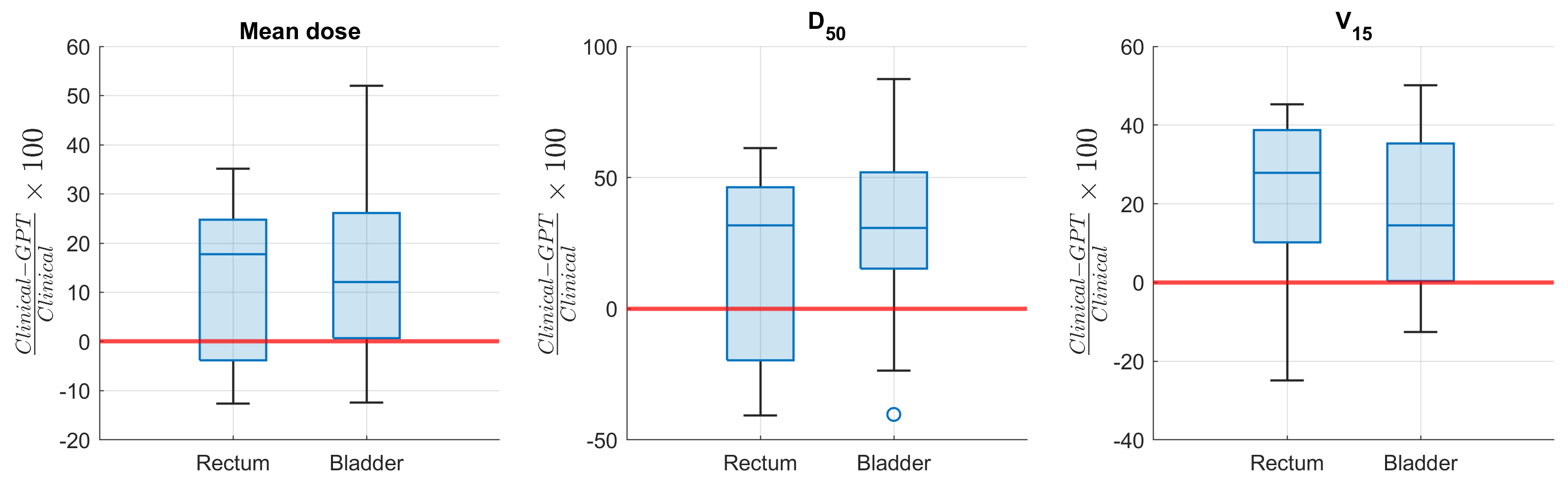}
        
        \label{fig:rel_prostate}
    \end{subfigure}
    \begin{subfigure}[]{\textwidth}
          \centering
           \caption{\textbf{Head \& neck patients}}
          \includegraphics[width=\textwidth]{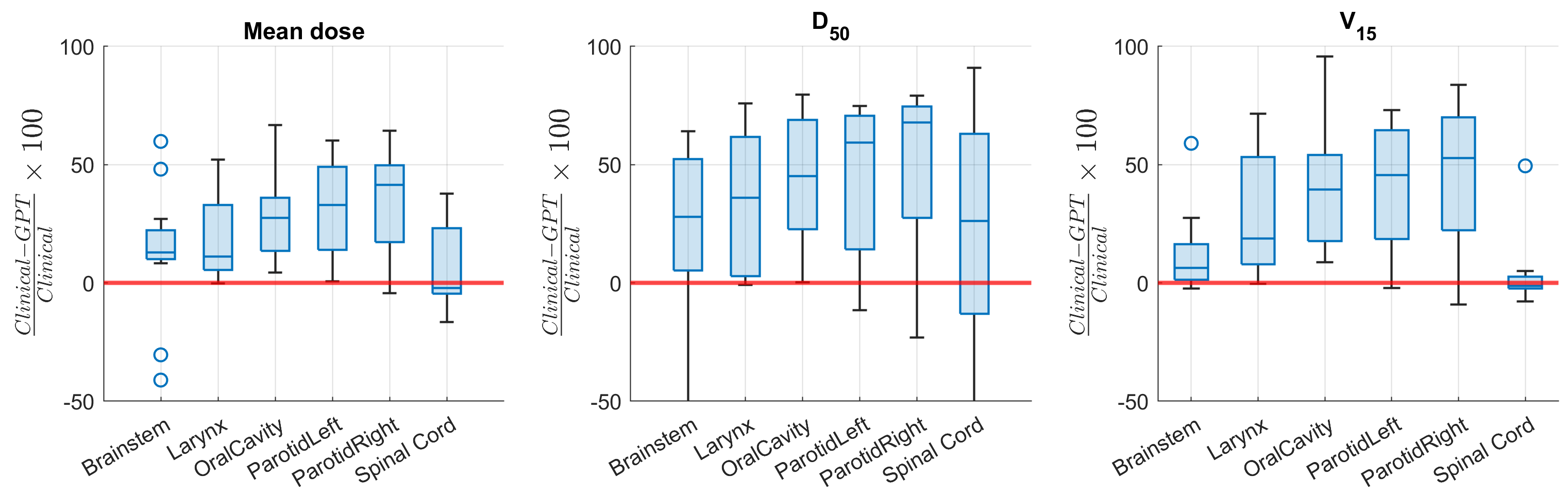}
         
          \label{fig:rel_hn}
    \end{subfigure}
    \caption{\textbf{Differences between GPT-RadPlan and clinical plans.} For every relevant OAR, each box displays the relative difference between clinical and GPT-RadPlan plans of all the patients in the prostate or head \& neck cohorts. Each panel shows the relative difference between different OAR metrics, including (left) the mean dose, (center) the $D_{50}$ and (right) the $V_{15}$. Positive values above the red line indicate that GPT-RadPlan plans result in lower metric values, which is preferred. }
    \label{fig:rel}
\end{figure*}

Due to the overlapping standard deviations noted in Table \ref{tab:oar} and the wide shaded regions in Figure \ref{fig:dvh}, GPT-RadPlan’s superiority over clinical plans in sparing OARs remains uncertain. The large standard deviations occur because the tumor targets in each patient are located in closer proximity to different OARs. For instance, two head \& neck patients each have a larger PTV overlapping either the left or right parotid gland, leading to a substantially higher dose received by the overlapping gland. We demonstrate this and prove GPT-RadPlan's superiority in Figure \ref{fig:rel}, where the difference in dose metrics between GPT-RadPlan and clinical plans is calculated individually for each patient \rebutal{in the prostate (Figure \ref{fig:rel_prostate}) and head \& neck (Figure \ref{fig:rel_hn})cohorts}. The data show that GPT-RadPlan's metrics are lower, as the boxes generally sit above the red line, confirming that GPT-RadPlan is capable of achieving plans that deliver less dose on the same patient. In general, the GPT-RadPlan plans \rebutal{reduce OAR doses while mantaining coverage} in 82\% of prostate patients, achieving an average reduction \rebutal{5 Gy decrease in mean dose} and 15\% overall in OAR metrics as shown in \ref{fig:rel_prostate}. Likewise, GPT-RadPlan \rebutal{improves OAR sparing} in 75\% of the head \& neck cases, reducing the mean dose by \rebutal{4-6 Gy} or 10-15\% for most of the OARs.

\rebutal{Regarding the non-clinical baselines, GPT-RadPlan also yields better plans than the BO and Autoplan, as demonstrated by the lower OAR metrics in Table \ref{tab:oar}. In particular, Autoplan appears to perform better than BO, similar to clinical planners and closer to GPT-RadPlan. For this reason, we have included additional results in Appendix \ref{app:ap}. As shown in \ref{fig:avg_dvh_ap}, although Autoplan often yields plans that are slightly worse than GPT-RadPlan's results, often with higher OAR doses except for select organs such as the femoral heads or the brainstem. The relative differences in Figure \ref{fig:rel_ap} in the Appendix, showing overall values that fall above the red line, further confirm GPT-RadPlan's superiority.}

\begin{figure*}[]
    \centering
    \begin{subfigure}[]{\textwidth}
        \centering
        \includegraphics[width=\textwidth]{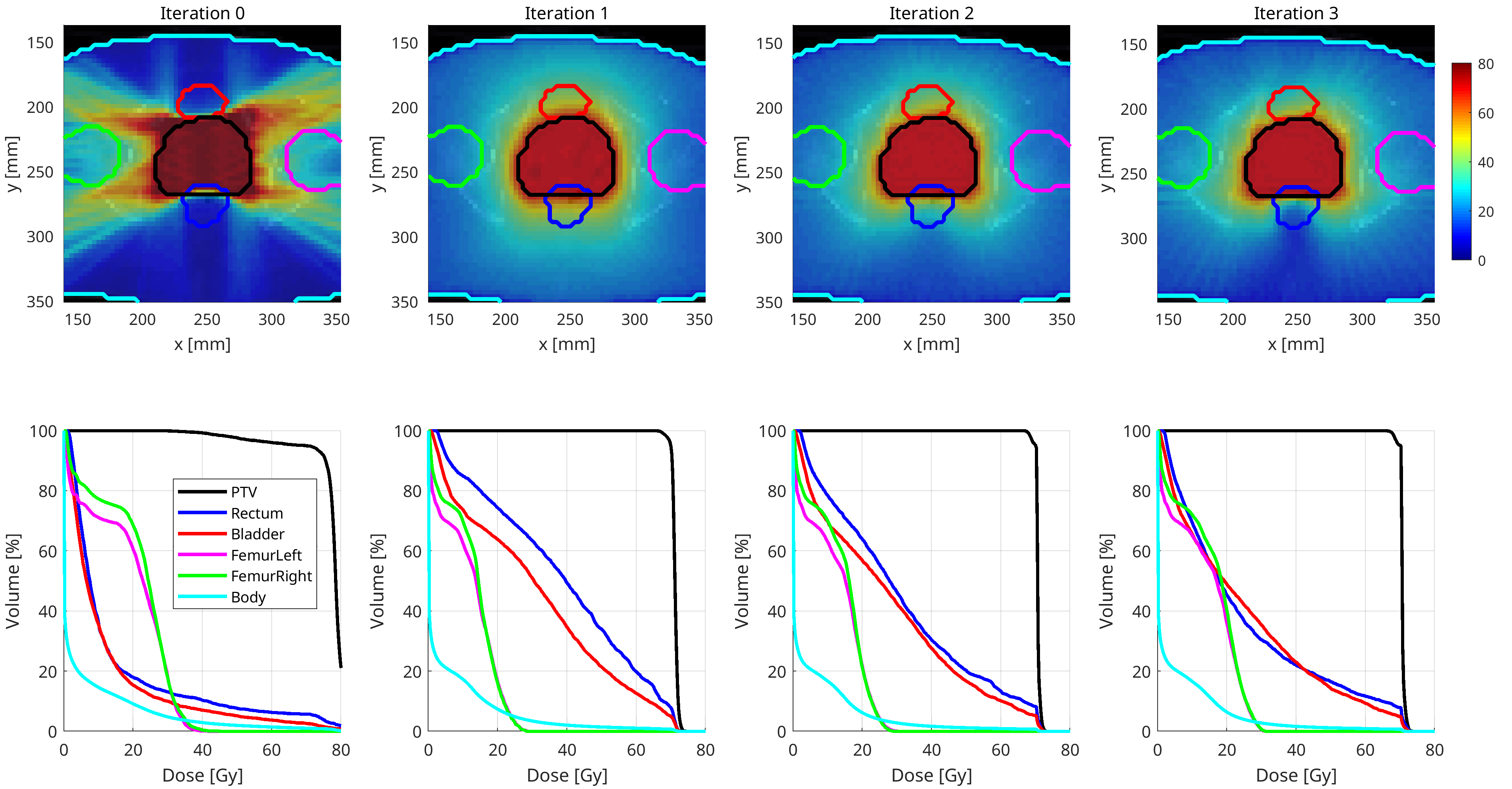}
        \caption{Prostate case}
        \label{fig:traj_prostate}
    \end{subfigure}
    \begin{subfigure}[]{\textwidth}
        \centering
        \includegraphics[width=\textwidth]{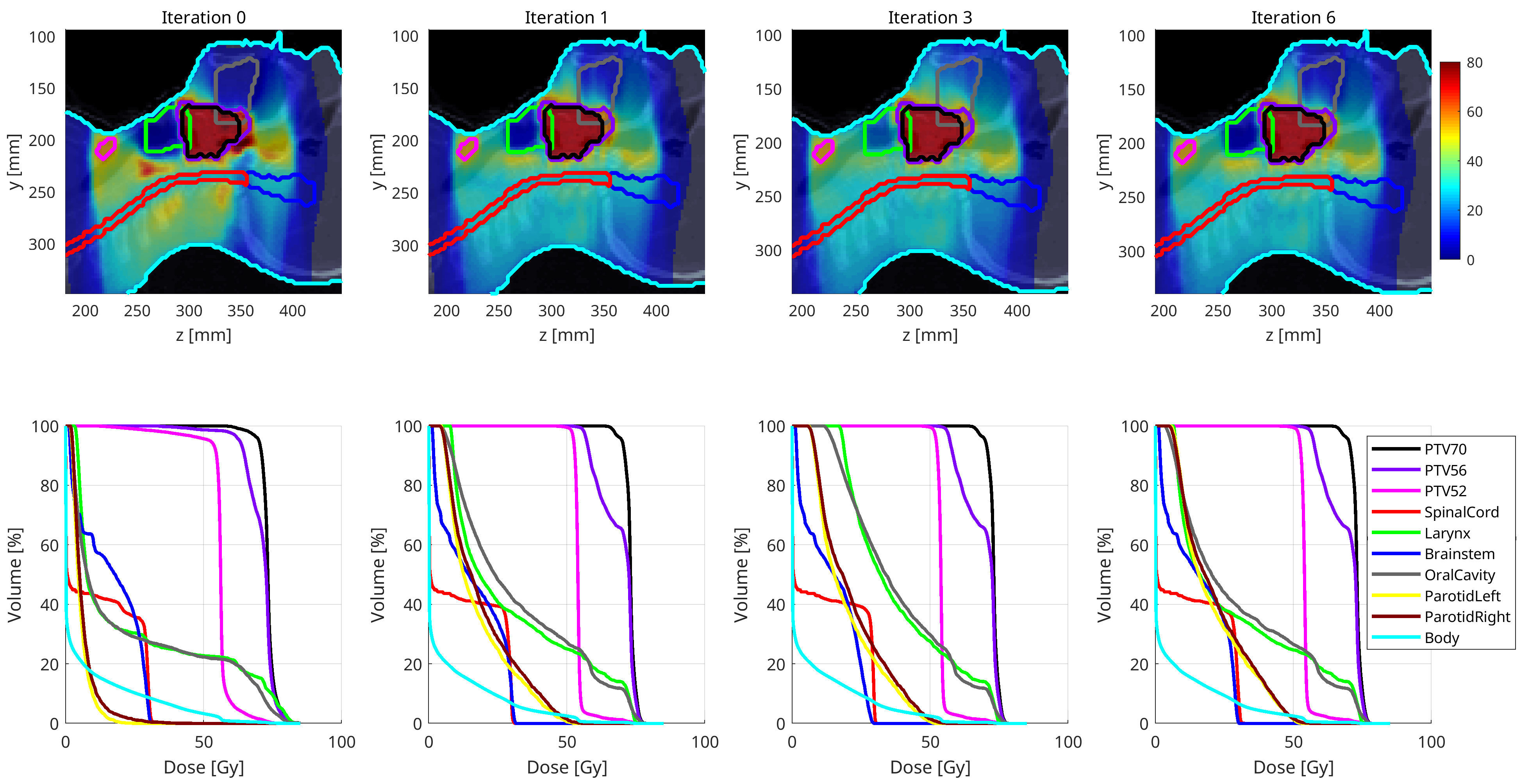}
        \caption{Head \& neck case}
        \label{fig:traj_hn}
    \end{subfigure}
    \caption{\textbf{Planning trajectories}. From left to right, each column shows the evolution of the dose distribution (2D slice centered at the iso-center) and the corresponding DVH. The top figure (a) shows one of the prostate cases requiring only 3 steps to obtain a treatment plan that meets the clinical protocol requirements, while (b) depicts the trajectory for a head \& neck case requiring 6 optimization steps. As the iteration number increases, GPT-RadPlan first ensures homogeneous PTV coverage, subsequently reducing the dose delivered to the OARs.}
    \label{fig:traj}
\end{figure*}

To investigate how GPT-RadPlan acts over the multiple iterations, Figure \ref{fig:traj} compares the dose distributions and DVHs for a prostate and head \& neck patient at different iteration steps. Figure \ref{fig:traj} depicts such \textit{planning trajectories} for one of the prostate patients, where GPT-RadPlan first ensures homogeneous PTV coverage (uniform red color) and subsequently reduces the dose delivered to the rectum and bladder (blue and red contours, respectively). Likewise, Figure \ref{fig:traj} demonstrates the same behavior for a head \& neck patient, where GPT-RadPlan first ensures target coverage (black, purple, and magenta contours), and then progressively minimizes the dose delivered to the oral cavity and larynx (grey and green contours), as well as the overall body dose. Most importantly, \rebutal{GPT-RadPlan takes 3-6 iterations to obtain the final plan (around 2-3 hours without human supervision needed), which is comparable to human planners and} lower than the 7-10 iterations for Autoplan and more than 50 iterations for the BO baseline.

\section{Discussion}
The proposed GPT-RadPlan represents an advancement in the field of radiation oncology, offering a fully automated treatment planning framework that leverages the capabilities of state-of-the-art multi-modal LLMs, such as OpenAI GPT-4V. Its ability to operate without the need for domain-specific datasets or extensive training potentially facilitates integrating GPT-RadPlan into various clinical workflows without requiring any adjustment to current clinical software and optimization algorithms. Notably, the code we use is identical for both prostate and head \& neck patients, with the only variation being in the input configuration file, which specifies the location of reference cases, the physician’s intent in text form, and a few other machine parameters. Moreover, being clinical software agnostic, GPT-RadPlan has the potential to be readily adopted across diverse disease sites and institutions, underscoring its broad applicability. To our knowledge, this is the first time that an automated treatment planning system based on LLMs has been proposed and demonstrated to be feasible for direct application to clinical environments without requiring any extensive model training. 

GPT-RadPlan has two advantages over any other learning-based method. First, it does not require a large dataset or time-consuming training, leveraging LLM's in-context learning abilities instead with only a handful of approved clinical plans. Second, it uses all the available information to evaluate and improve plans, including dose distribution, DVH, clinical preferences (as text) and quantitative metrics. Moreover, users can potentially specify clinical preferences for a specific patient using human language \rebutal{(though we did not do it in our experiments}, making GPT-RadPlan a versatile tool that can potentially be adapted to multiple disease sites and requirements using the same code and exclusively changing the prompt. As such, GPT-RadPlan can optimize plans to directly satisfy prescribed clinical criteria as defined by radiation oncologists in the physician's intent, therefore avoiding the use of proxy metrics or crafted wish-lists. The reduced need for human interventions during planning further enhances clinical workflow and improves the allocation of human resources, freeing radiation oncologists and dosimetrists to focus on other critical aspects of patient care.

GPT-RadPlan also offers several advantages in comparison to previous RL auto-planning methods \cite{shenOperatingTreatmentPlanning2020}. In general, deep RL algorithms approximate policy functions and take actions in a 'black-box' manner, where decisions are hardly interpretable. In contrast, GPT-RadPlan reflects on its previous results, suggesting improvements and acting based on these reflections. As a result, GPT-RadPlan actions (i.e., weight adjustments) always have an explanation. Moreover, compared to the previous RL studies requiring an entire week to train an RL agent for a specific disease site and protocol, GPT-RadPlan massively reduces computation times, being directly applicable after only minor prompt modifications. Lastly, unlike RL agents requiring a reward that hardly captures all clinical objectives and human preferences, the proposed GPT-RadPlan agents directly evaluate plans based on the entire prescribed protocol fully capturing the physician's intent.

As demonstrated in the results, the plans generated by GPT-RadPlan uniformly cover PTV targets and spare OARs, providing a high level of consistency and quality that should contribute to better patient outcomes. After multiple iterations, GPT-RadPlan can provide plans that satisfy all the criteria in the clinical protocol for prostate cases, matching the PTV coverage requirements while always irradiating OARs with lower mean doses and DVH constraints than those indicated in the protocol. For head \& neck plans, GPT-RadPlan’s plans cannot satisfy all the clinical requirements simultaneously, as is also often observed in clinical practice. This involves conflicting objectives, such as sparing two different organs or prioritizing the coverage of the PTVs rather than sparing a single OAR (e.g., the larynx). Since the \rebutal{AI} planner prioritizes dose objectives based on the order specified in the prompt, GPT-RadPlan could, in principle, allow for the exploration of the Pareto frontier by easily modifying these prompt priorities.

\paragraph{Future steps} As possible next steps, instead of providing a few manually selected patients as a reference, GPT-RadPlan could be coupled to retrieval-augmented generation (RAG) that retrieves the most similar plans in a database to be used for assistance \cite{huangLearningImageRepresentations2023,liu2022deep}. Future work could also explore GPT-RadPlan applications to proton treatments in the presence of setup, range or anatomical uncertainties \cite{pastor-serranoHowShouldWe2021,pastor-serranoProbabilisticDeepLearning2023a}, expanding the model capabilities \rebutal{as a treament planninf copilot to automatically select} robustness settings \rebutal{or the anatomies or scenarios included during robust treatment planning}. This work proposed novel method to automate treatment planning using pre-trained LLMs that shows promising results in just two disease sites, introducing a new line of research that explores the advantages of applying LLMs to assist human clinicians. To demonstrate generalization across disease sites, future work should focus on the clinical evaluation of GPT-RadPlan in more diverse clinical settings and benchmarks.

\paragraph{Limitations}
\rebutal{This paper presents the novel idea of using LLMs as a copilot during treatment planning to mimic human planners decisions, but the TPS used to obtain clinical plans (Eclipse) is different from the one that GPT-RadPlan is coupled to. Although the objectives and problem formulation are the same, subsequent studies must compare GPT-RadPlan with human clinical planners using the same TPS.}

One limitation of GPT-RadPlan is its lack of direct alignment \rebutal{with physicians' preferences} through model fine-tuning \rebutal{as we directly use the GPT-4V, which is not specifically trained for the task}.  While it is true that OpenAI trained their models on large-scale datasets that may include detailed clinical information, we do not supply any specific information about optimal planning choices for the patients. The model can only be made aware of clinical preferences via the physician's intent in the prompt, which limits the ability to specify trade-off preferences that could significantly differ between areas, e.g., the larynx and oropharynx. Subsequent studies could further improve model performance by manually collecting radiation oncologists' plan evaluations and using these to fine-tune the LLMs and calibrating the LLMs~\cite{liu2024calibrating,liu2022avoiding,liu2022deep}, or using RT fine-tuned models to assist in the evaluation task \cite{liuRadOncGPTLargeLanguage2023}. Due to the dependence on input clinical protocols and reference cases, ambiguous or inadequate input could result in GPT-RadPlan generating suboptimal or potentially unsafe treatment plans. Furthermore, the use of non-representative reference cases may introduce bias, limiting the generalization of the model across diverse patient populations or institutions. While GPT-RadPlan provides interpretable feedback for parameter adjustments, its underlying decision-making remains partially opaque, which could hinder clinicians' ability to fully understand or trust its recommendations.

To mitigate these risks, continuous physician supervision is crucial to ensure adherence to safety and quality standards. Regular calibration and testing of GPT-RadPlan using real-world feedback can help maintain its reliability and adapt it to diverse clinical settings. In addition, incorporating automated validation mechanisms to flag outlier plans or deviations from clinical guidelines can serve as a robust safety net. 

Future efforts should focus on fine-tuning GPT-RadPlan with localized datasets to enhance its adaptability, developing explainability frameworks for greater transparency, and integrating uncertainty quantification methods to identify high-risk predictions. These steps will further improve the reliability and clinical utility of GPT-RadPlan.

\rebutal{Finally, a limitation of GPT-RadPlan is its inability to match the Pareto optimality guarantees of multicriteria optimization methods \cite{breedveldNovelApproachMulticriteria2007, breedveldICycleIntegratedMulticriterial2012a}, which generate a Pareto optimal set of plans where no single objective (like tumor coverage or organ sparing) can be improved without compromising another. Trial-and-error methods such GPT-RadPlan do not explore the Pareto front but rather converge to a solution that is clinically acceptable based on physician intent and specified preferences. While, as demonstrated in the results, this final plan can perform well, GPT-RadPlan does not provide Pareto optimality guarantees or a clear view of the planning trade-offs. }

\section{Conclusion}
In this study, we present a novel GPT-RadPlan framework to enhance the collaborative performance of radiotherapy treatment planning workflow based on GPT-4V. Building upon the commonsense reasoning capabilities and domain knowledge on radiation oncology exhibited by GPT-4V, without any further fine-tuning or training on the GPT-4V, we effectively augment the planning and coordination abilities among multi-modal LLM agents through context-aware reasoning mechanisms and comprehensive feedback mechanisms, facilitating continuous interaction between the modules like planner and evaluator. GPT-RadPlan demonstrates remarkable performance in prostate and head \& neck planning. Notably, it exhibits exceptional capabilities in learning from historical trajectories and mistakes in the past. We believe that with further enhancements in multi-modal LLMs, using these for treatment planning will provide new opportunities for further advancement.

\bibliographystyle{unsrt}
\bibliography{llm_opt}

\newpage
\appendix
\rebutal{
\section{Prompts}
\label{sec:tp_prompts}
Radiotherapy treatment planning involves an evaluator that evaluates the inverse optimized plan, and a planner that produces the parameters for the next iteration. We adopt multimodal LLMs to evaluate and produce new parameters by prompting them to perform the corresponding tasks. Specifically, for the evaluation, a multimodal LLM is used to compare each protocol with the current plan and produce the final assessment. Similarly, we prompt a multimodal LLM to produce new parameters based on the plan in the current iteration as well as feedback from the evaluator. 
\label{tp_prompts}
\begin{figure}[!t]
\begin{apxtcolorbox}[Evaluator prompt]
Please act as an impartial and objective professional radiation oncologist. Your job is to evaluate the quality of radiation therapy treatment plans based on their dose volume histogram. You should consider the following protocol: 

<Clinical Protocol>

Here is the dose-volume histograms of the candidate plans for evaluation; each entry in the dose-volume histograms (DVH) table indicates the percentage of volume receiving a dose higher than a certain Gy (specified in the first column). 

<DVH table>

We also provide the statistics of the above DVH table for ease of evaluation. 

<DVH statistics>

You should also evaluate the plan based on the dose distribution, considering the anatomy of the patient. For your evaluation, you should consider the following factors:

\begin{itemize}
    \item Homogeneity and conformity: A good plan should have a conformal and homogeneous dose (red in the image) within the PTV. There should be no hot spots or cold spots.
    \item Dose spillage. A good plan should NOT have high (red) dose spillage outside of the PTV. The does distribution must show a sharp gradient outside the PTV/target.
\end{itemize}

<Images>

Now, based on the protocol, the images and the DVH, please evaluate the plans. Avoid any positional biases and ensure that the order in which the responses are presented does not influence your decision. Be as objective as possible. Your answer must: 
\begin{enumerate}
    \item (Evidence) Extract the corresponding entries from the DVH table based on the factors to consider.
    \item (Interpretation) Based on the extracted entries from DVH table, interpret them as: \texttt{\{percentage\}\%} of volume encompassing the dose more than \texttt{\{dose\}} Gy for \texttt{\{organ\}}.
    \item (Analysis) Combine the interpretation and the factors, analyze whether each of the numerical factors are met.
\end{enumerate}
Based on the analysis on DVH tables, you need to produce a final evaluation. When all factors are satisfied, you should always answer no improvement is needed. If improvements are required, please provide suggestions on where to improve. Please ensure to follow the format below to return the evaluation results:\\
\\
<FINAL> Decision: [The plan does/doesn't need to be improved.] Reasons: [list all factors that are not satisfied with detailed reasons, e.g. protocol X on (PTV/rectum/bladder/fh/body) is not satisfied because ...] </FINAL>\\
\\
The final answer in the end must strictly follow the format above.

\end{apxtcolorbox}
\end{figure}

\begin{figure}[!t]
\begin{apxtcolorbox}[Planner prompt]
Please act as a professional radiation oncologist who is impartial and objective. Your job is to provide a set of parameters to optimize the treatment plan. Treatment planning involves a balancing act between competing clinical goals. The desirable dose 
distribution is determined by an optimization algorithm minimizing a cost function with multiple competing objectives. Each objective is the difference between the current dose and the prescribed dose of a certain organ (PTV, brainstem, etc.). The parameters decide the importance of each objective. \\

You are provided with a set of good reference plans, the DVH, and their corresponding parameters, to learn what the parameters of good plans should look like. \\

Reference treatment plan DVH: \\
<Reference plan DVH>\\
Reference treatment plan parameters: \\
<Reference plan parameters>\\

In the following, you are provided with the dose volume histograms (DVHs) of the treatment plans in the previous iterations as well as the parameter weights used to generate the plans.\\

Here are suggestions from the evaluator on improving the historical treatment plan: \\
<suggestions on improving the plan>\\

Now, based on the historical treatment plan as well as the suggestion on what to improve, you need to generate a new set of 
parameters to rebalance the objectives of the optimization: \\

Please ensure to conclude with the following one-line dictionary format (without name): \\

    {{"PTV70": <weight>, "PTV56": <weight>, "PTV52": <weight>, "Brainstem": <weight>, "SpinalCord": <weight>, "ParotidLeft": <weight>, "ParotidRight": <weight>, 
    "OralCavity": <weight>, "Larynx": <weight>, "Body": <weight>}}\\

The final answer in the end must strictly follow this format in one single line to get a valid Python dictionary that I can directly parse it using Python.

\end{apxtcolorbox}
\end{figure}

\section{Additional results}
In this section, we present metrics and figures to further evaluate the OAR sparing capabilities of GPT-RadPlan with respect to clinical plans (Section \ref{app:cli}). We also include additional plots comparing the proposed method to the non-clinical Autoplan baseline in Section \ref{app:ap}.

\subsection{Clinical plan comparison}
\label{app:cli}
As observed in Table \ref{tab:ptv}, GPT-RadPlan delivers a lower mean dose to the PTV that is closer to prescription, especially in prostate treatments. To discard lower total average doses as a potential cause of observing lower OAR mean doses and metrics, we have rescaled the GPT-RadPlan plan doses so that the mean PTV dose matches that of clinical plans. Based on these rescaled doses, Table \ref{tab:oar_norm} shows the recalculated OAR metrics. These are the same metrics and calculations as in Table \ref{tab:oar}.  With an average 4 Gy difference for prostate plans and  6 Gy difference for head \& neck plans, these results confirm GPT-RadPlan's OAR sparing capabilities. 

\begin{table}[h]
\caption{\textbf{OAR dose metrics after dose scaling}. Multiple dose metrics capturing OAR sparing in prostate and head \& neck plans are presented, after scaling the GPT-RadPlan doses so that the mean PTV doses match those of the clinical plans. These include the mean dose, the $D_5$ and $D_{50}$ (maximum common dose that 5\% or 50\% of the volume receives), and the $V_{15}$ and $V_{30}$ (representing which percentage of the volume receives 15 and 30 Gy, respectively). For all the metrics, we include the average values across plans and the standard deviation in brackets. In all cases, lower values are preferred.}
\footnotesize
    \begin{tabular}{@{}lllllll@{}}
    \toprule
    \textbf{Organ} & \textbf{Method} & \textbf{Mean dose [Gy]} & \textbf{$\textbf{D}_{5}$ [Gy]} & \textbf{$\textbf{D}_{50}$ [Gy]} & \textbf{$\textbf{V}_{15}$ [\%]} & \textbf{$\textbf{V}_{30}$ [\%]} \\ \midrule
    
    \multicolumn{7}{c}{Prostate plans} \\\midrule
    \multirow{2}{*}{Bladder} & Clinical & 24.48 (9.30) & 62.14 (10.79) & 18.87 (12.12) & 60.85 (29.52) & 34.45 (15.28) \\
                             & GPT-RadPlan & 17.41 (8.34) & 61.42 (14.77) & 9.41 (8.85) & 36.41 (17.52) & 23.13 (13.46) \\ \midrule
                             
    \multirow{2}{*}{Rectum} & Clinical & 26.45 (5.64) & 63.64 (8.82) & 22.20 (5.71) & 71.91 (14.92) & 32.87 (10.54) \\
                            & GPT-RadPlan & 22.27 (6.68) & 60.18 (14.13) & 17.44 (8.06) & 51.52 (20.39) & 28.21 (16.64) \\ \midrule

    \multicolumn{7}{c}{Head \& neck plans} \\\midrule
    \multirow{2}{*}{Brainstem} & Clinical & 12.17 (4.29) & 30.95 (4.78) & 8.34 (5.75) & 33.36 (14.10) & 10.03 (11.03) \\
                               & GPT-RadPlan & 10.8 (4.82) & 26.57 (4.65) & 7.98 (7.41) & 32.54 (15.84) & 0.03 (0.1) \\ \midrule
                               
    \multirow{2}{*}{Larynx} & Clinical & 42.80 (11.03) & 69.95 (4.30) & 39.85 (14.87) & 97.44 (3.24) & 64.06 (18.18) \\
                            & GPT-RadPlan & 33.78 (14.36) & 69.82 (5.77) & 26.71 (21.6) & 63.42 (26.83) & 42.04 (24.98) \\ \midrule
                            
    \multirow{2}{*}{Oral Cavity} & Clinical & 34.86 (4.15) & 68.07 (8.83) & 33.62 (12.29) & 92.96 (5.28) & 50.00 (11.01) \\
                                 & GPT-RadPlan & 25.12 (8.45) & 65.58 (14.99) & 18.04 (8.26) & 53.74 (25.5) & 27.59 (13.65) \\ \midrule
                                 
    \multirow{2}{*}{Parotid Left} & Clinical & 35.10 (13.93) & 63.18 (10.85) & 33.50 (21.21) & 71.64 (20.97) & 52.80 (24.12) \\
                                  & GPT-RadPlan & 23.99 (12.54) & 57.08 (15.96) & 19.51 (18.42) & 42.5 (20.52) & 28.18 (20.87) \\ \midrule
                                  
    \multirow{2}{*}{Parotid Right} & Clinical & 34.32 (10.67) & 61.69 (11.30) & 32.22 (13.83) & 79.00 (21.05) & 52.65 (20.65) \\
                                   & GPT-RadPlan & 21.41 (7.74) & 57.8 (15.47) & 13.51 (7.41) & 41.36 (19.55) & 25.33 (13.72) \\ \midrule
                                   
    \multirow{2}{*}{Spinal Cord} & Clinical & 17.36 (6.32) & 33.88 (5.15) & 18.84 (13.24) & 55.32 (13.69) & 24.83 (25.69) \\
                                 & GPT-RadPlan & 16.07 (3.93) & 29.77 (0.3) & 17.56 (12.37) & 55.12 (14.1) & 1.33 (3.22) \\ \bottomrule
    \end{tabular}
    \centering
    \label{tab:oar_norm}
\end{table}

Furthermore, to help visualize the individual plan-by-plan differences between GPT-RadPlan and clinical plans from Figure \ref{fig:rel}, we have included additional Figure \ref{fig:diffb_prostate} and Figure \ref{fig:diffb_hn}, where each bar shows the same relative error for every OAR and plan in the prostate and head \& neck cohorts, respectively. Positive values indicate higher clinical plan metrics. Since the majority of the bars show positive values, we can further confirm that GPT-RadPlan achieves greater OAR sparing in both prostate and head \& neck plans.

\begin{figure*}[h]
    \centering
    \includegraphics[width=\textwidth]{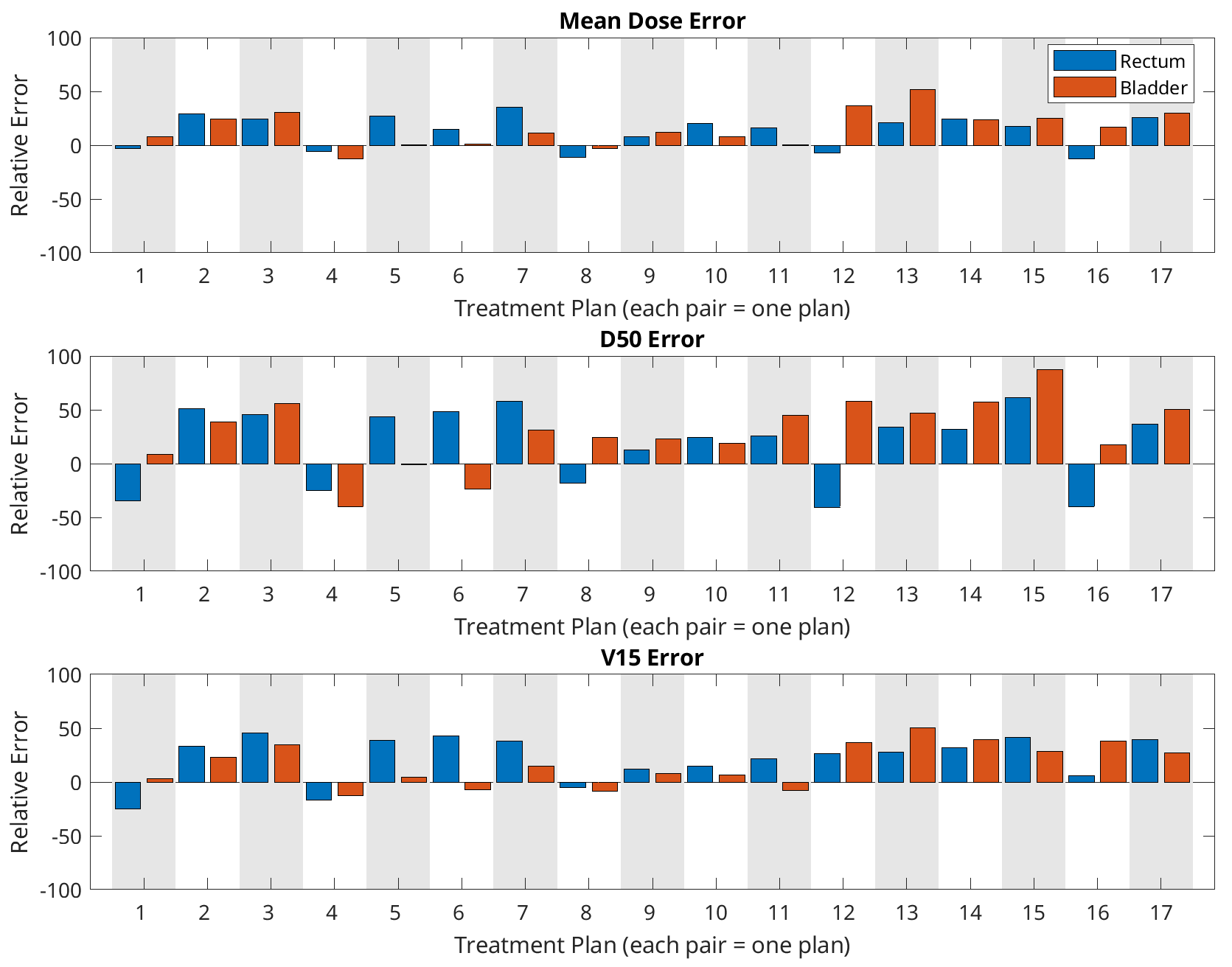}
    \caption{\textbf{Relative difference between GPT-RadPlan and clinical prostate plans.} For every relevant OAR, each plot depicts the relative difference between clinical and GPT-RadPlan plans for 3 different metrics (mean dose, $D_{50}$ and $V_{15}$. Each white and gray block (each pair of bars) shows the results for a separate plan. Positive values indicate GPT-RadPlan superiority over clinical plans.}
    \label{fig:diffb_prostate}
\end{figure*}

\begin{figure*}[h]
    \centering
    \includegraphics[width=\textwidth]{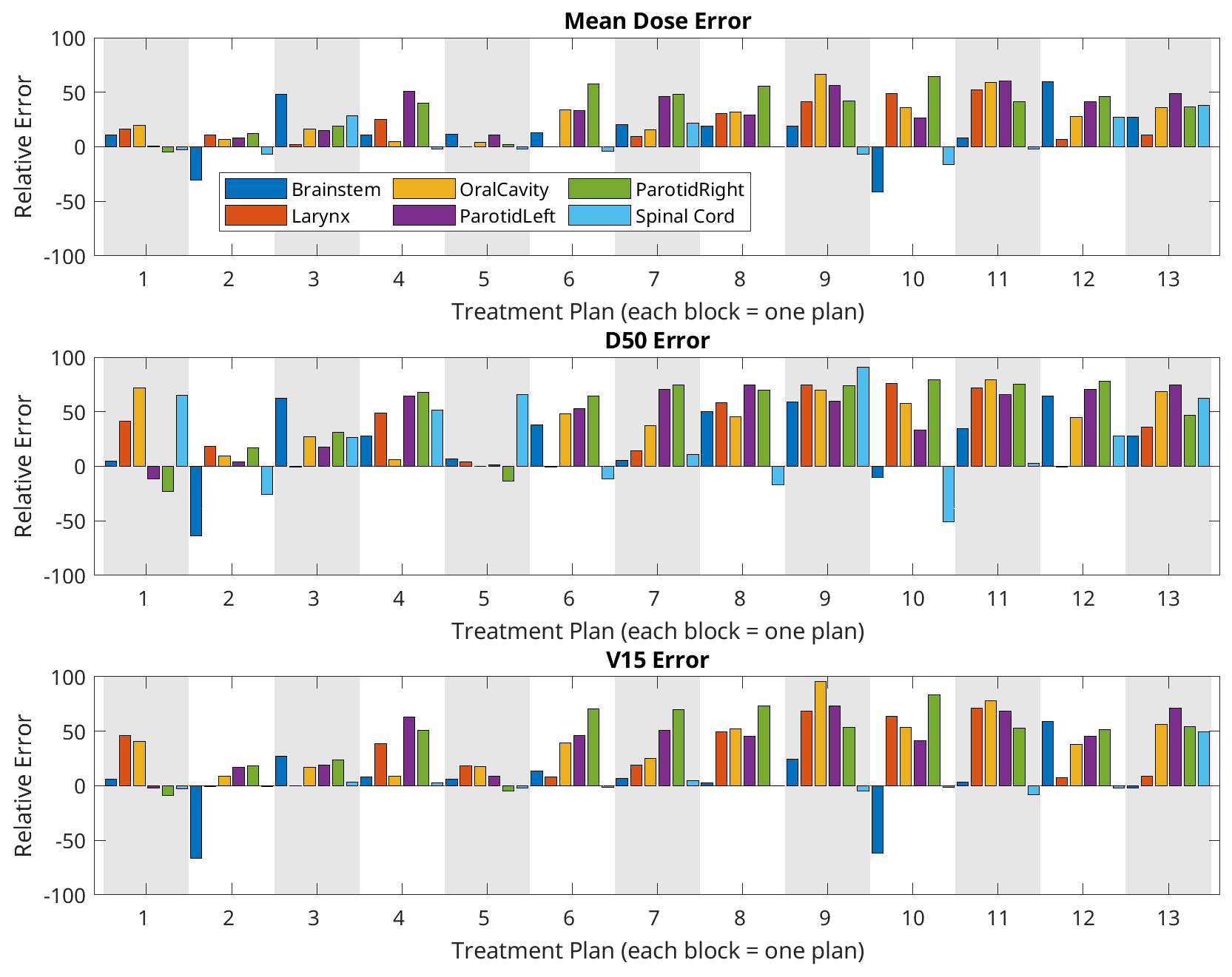}
    \caption{\textbf{Relative difference between GPT-RadPlan and clinical heand and neck plans.} For every relevant OAR, each plot depicts the relative difference between clinical and GPT-RadPlan plans for 3 different metrics (mean dose, $D_{50}$ and $V_{15}$. Each white and gray block with 6 bars shows the results for each of the plans individually. Positive values indicate GPT-RadPlan superiority over clinical plans.}
    \label{fig:diffb_hn}
\end{figure*}

\subsection{Autoplan comparison}
\label{app:ap}

To further demonstrate GPT-RadPlan's superiority to Autoplan, we have included the same plots and comparison used for comparison with clinical plans. Figure \ref{fig:avg_dvh_ap} compares the average DVH curves for both planning methodologies, which indicate a slightly more homogeneous PTV coverage in GPT-RadPlan plans, and overall improved OAR sparing, except for the femoral heads in prostate plans. For head \& neck plans, Autoplan results in lower volumes of the brainstem and spinal cord receiving doses in the 10-20Gy range, although the maximum doses are ultimately higher. 

Figure \ref{fig:rel_ap} further compares individual relative differences between plans, demonstrating that GPT-RadPlan outperforms Autoplan in most situations as shown by the positive values above the red line, except for the brainstem and spinal cord as shown in Figure \ref{fig:rel_hn_ap}.

\begin{figure}[h]
    \centering
    \includegraphics[width=\textwidth]{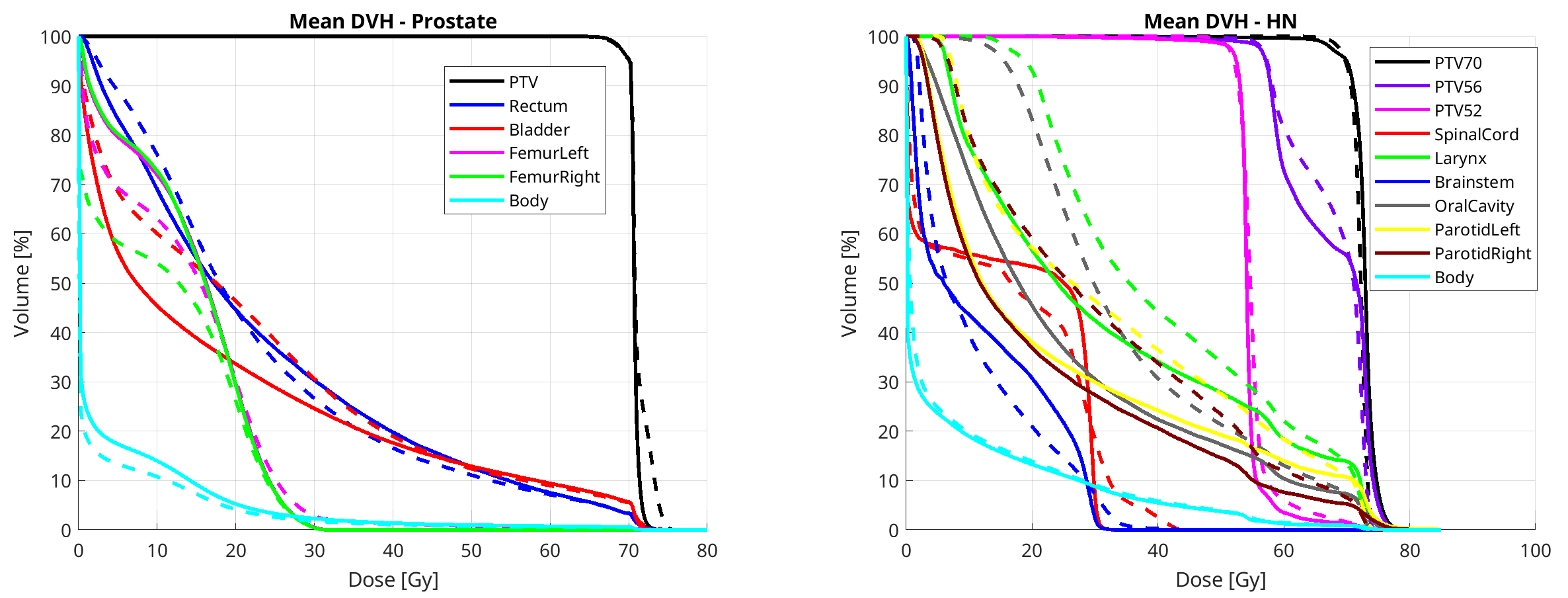}
    \caption{\textbf{Average DVH comparison between GPT-RadPlan and Autoplan}. Visual comparison of the average DVH lines across all prostate (left) and lung (right) patients. Solid lines represent GPT-RadPlan plans, while dashed lines indicate Autoplan plans.}
    \label{fig:avg_dvh_ap}
\end{figure}

\begin{figure*}[h]
    \centering
    \begin{subfigure}[]{\textwidth}
        \centering
        \caption{\textbf{Prostate patients}}
        \includegraphics[width=\textwidth]{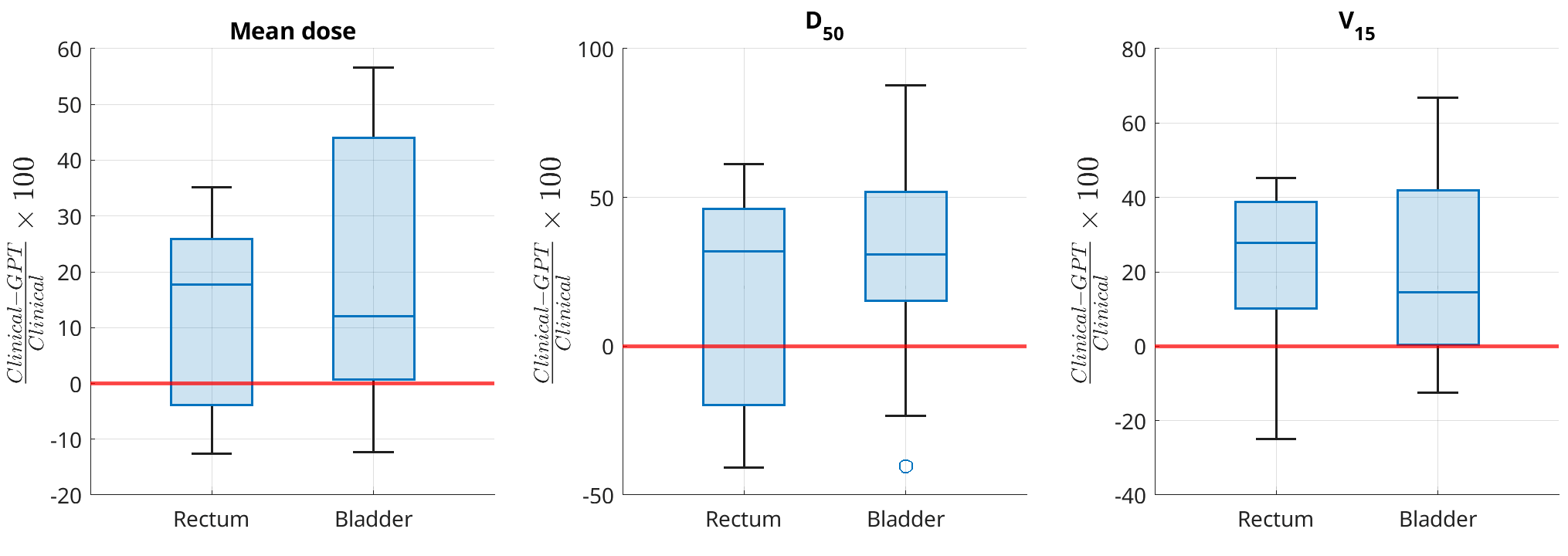}
        
        \label{fig:rel_prostate_ap}
    \end{subfigure}
    \begin{subfigure}[]{\textwidth}
          \centering
           \caption{\textbf{Head \& neck patients}}
          \includegraphics[width=\textwidth]{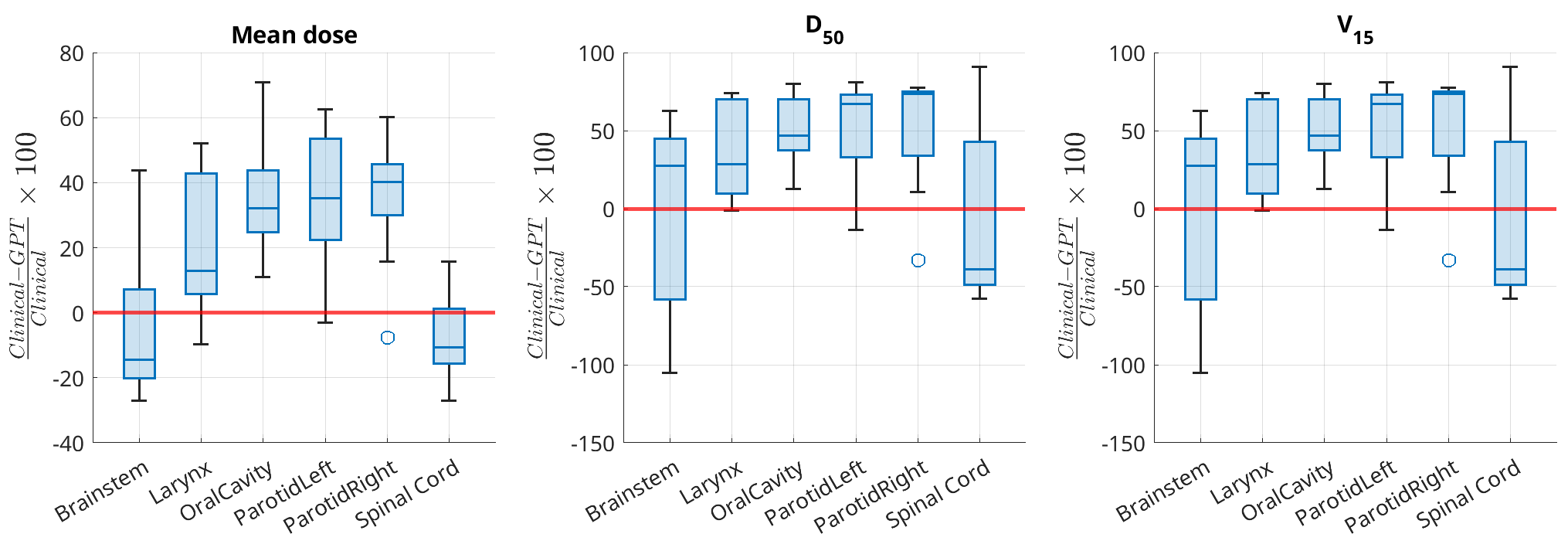}
         
          \label{fig:rel_hn_ap}
    \end{subfigure}
    \caption{\textbf{Differences between GPT-RadPlan and Autoplan plans.} For every relevant OAR, each box displays the relative difference between plans obtained using the Autoplan and GPT-RadPlan algorithms. Each panel shows the relative difference between different OAR metrics, including (left) the mean dose, (center) the $D_{50}$ and (right) the $V_{15}$. Positive values above the red line indicate that GPT-RadPlan plans result in lower metric values, which is preferred. }
    \label{fig:rel_ap}
\end{figure*}

}

\end{document}